\documentclass[10pt]{article}

\usepackage{amsmath}
\usepackage{amssymb}
\usepackage{graphicx}
\usepackage{graphicx,epsfig}

\usepackage{cite}

\usepackage{color} 

\usepackage{setspace} 
\usepackage[labelfont=bf,labelsep=period,justification=raggedright]{caption}

\makeatletter
\renewcommand{\@biblabel}[1]{\quad#1.}
\makeatother

\date{}

\begin{document}

\begin{flushleft}
{\Large
\textbf{From modular to centralized organization of synchronization in functional areas of the cat cerebral cortex}
}
\\
Jes\'us G\'omez-Garde\~nes$^{1,2,\ast}$, 
Gorka Zamora-L\'opez$^{3}$, 
Yamir Moreno$^{1,4}$,
Alex Arenas$^{1,5}$
\\
\bf{1} Instituto de Biocomputaci\'on y F\'{\i}sica de Sistemas Complejos (BIFI), Universidad de Zaragoza, 50009 Zaragoza, Spain
\\
\bf{2} Departamento de Matem\'atica Aplicada, Universidad Rey Juan Carlos (ESCET), 95123 M\'ostoles (Madrid), Spain
\\
\bf{3} Interdisciplinary Center for Dynamics of Complex Systems,  University of Potsdam, D-14476 Potsdam, Germany
\\
\bf{4} Departamento de F\'{\i}sica Te\'orica, Universidad de Zaragoza, 50009 Zaragoza, Spain
\\
\bf{5} Departament d'Enginyeria Inform\'atica y Matematiques, Universitat Rovira i Virgili, 43007 Tarragona, Spain
\\
$\ast$ E-mail: gardenes@gmail.com
\end{flushleft}

\section*{Abstract}
\noindent Recent studies have pointed out the importance of transient
synchronization between widely distributed neural assemblies to
understand conscious perception. These neural assemblies form
intricate networks of neurons and synapses whose detailed map for
mammals is still unknown and far from our experimental
capabilities. Only in a few cases, for example the C. elegans, we know
the complete mapping of the neuronal tissue or its mesoscopic level of
description provided by cortical areas. Here we study the process of
transient and global synchronization using a simple model of
phase-coupled oscillators assigned to cortical areas in the cerebral
cat cortex. Our results highlight the impact of the topological
connectivity in the developing of synchronization, revealing a
transition in the synchronization organization that goes from a
modular decentralized coherence to a centralized synchronized regime
controlled by a few cortical areas forming a Rich-Club connectivity 
pattern.


\section*{Introduction}
Processing of information within the nervous system follows different
strategies and time-scales. Particular attributes of the sensory
stimuli are transduced into electrical signals of different
characteristics, e.g. regular or irregular spiking, the rate of
firing, etc. Further aspects of the information are ``encoded'' by
specialization of neurons, e.g. the color and orientation of a visual
stimulus will activate only a set of neurons and leave others
silent. For higher order processes such as feature binding and
association, the synchronization between neural assemblies plays a
crucial role~\cite{Engel,EngelReview,Fahle,Singer,Uhlhaas}. For example,
subliminal stimulation which is not consciously perceived, triggers a
similar cascade of activation in the sensory system but fails to
elicit a transient synchronization between distant cortical
regions~\cite{Melloni}.

The neurons comprising the nervous system form a complex network of
communications. To what extent this intricate architecture supports
the richness and complexity of the ongoing dynamical activity in the
brain is a fundamental question~\cite{Handbook}. A detailed map of the
neurons and their synapses in mammals is still unknown and far from
our experimental capabilities. Only in a few cases, for example the
nematode \emph{C. elegans}, we know the complete mapping of the
neuronal tissue. In the cases of macaque monkeys and cats a mesoscopic
level of description is known, composed of cortical areas and the
axonal projections between them. These areas are arranged into
modules which closely follow functional
subdivisions by modality~\cite{Scannell1995,Scannell1993,Hilgetag2000,Hilgetag2,Sporns,Hilgetag2004}. Two
cortical areas are more likely connected if both are involved in the
processing of the same modal information (visual, auditory, etc.)
Beyond this modular organization, some cortical areas are extensively
connected (referred as {\em hubs}) with projections to areas in all
modalities~\cite{SpornsHubs, Zamora2009}. For the corticocortical network 
of cats, these hubs are found to be densely
interconnected forming a {\em hidden} module~\cite{ZamoraIntegrator}, at
the top of the cortical hierarchy which might be responsible for the
integration of multisensory information. A core of cortical areas has also been
detected in estimates of human corticocortical connectivity by 
Diffusion Spectrum Imaging~\cite{HagmannHubs}. Following the above
discussion that synchronization plays a major role in the processing
of high level information, it would be important to analyze the
synchronization behavior of these networks in relation to their
modular and hierarchical organization. To this end, we simulate the
corticocortical network of the cat by non-identical phase oscillators
and we follow the evolution of their synchronization from local to
global.

The study of synchronization phenomena is a useful tool to analyze the
substrate of complex networks. The dynamical patterns under different
parameters unveil features of the underlying microscopic and
mesoscopic organization~\cite{SyncRev}. In particular, recent studies
highlight the impact that the topological properties such as the
degree heterogeneity, the small-world effect and the modular structure
have on the path followed from local to global synchronization
\cite{Alex-syncom,PRL,PRE,SyncRev}.

\begin{figure}[!ht]
\begin{center}
\epsfig{file=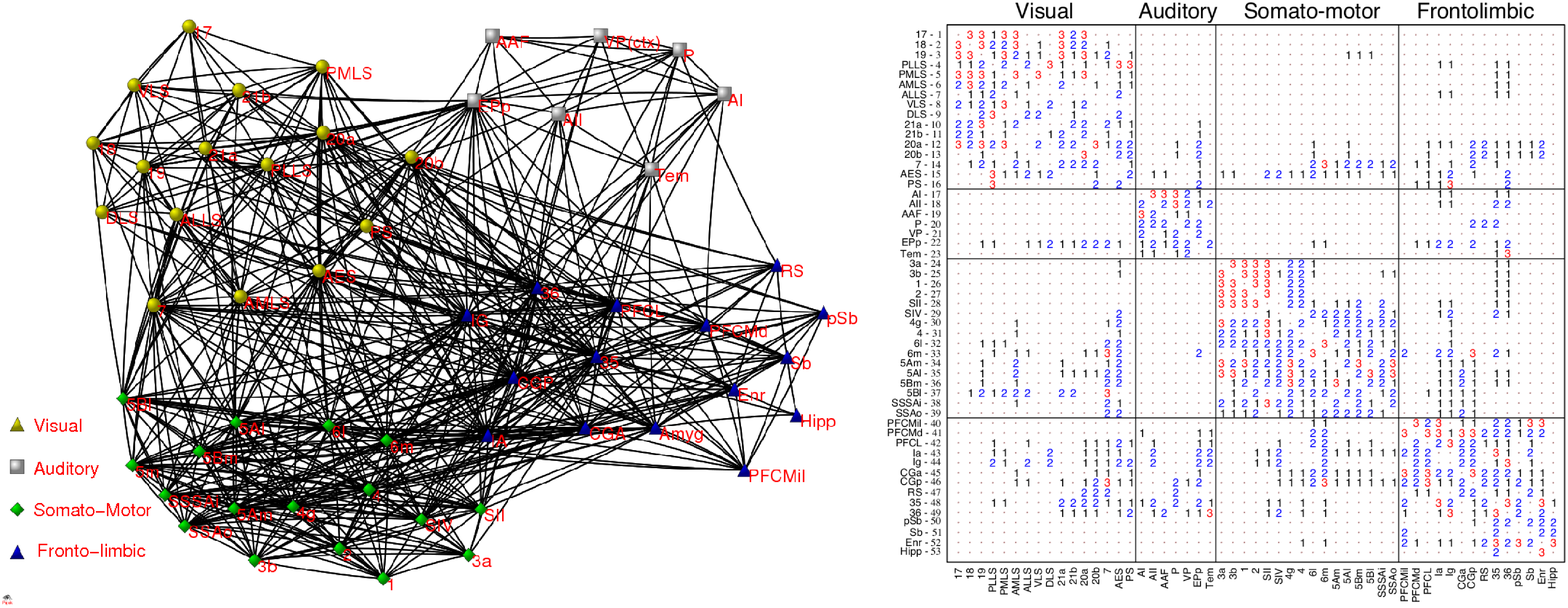,width=1.00\columnwidth,angle=-0,clip=1}
\end{center}
\caption{{\bf The brain cortical network of the cat}. On the left we
  show the topology of the nodes (areas) and links (axon
  interconnections) between them. On the right the weighted adjacency
  matrix is shown. The weight of the links denote the axon density
  between two connected areas. Besides the matrix shows the partition
  of the network into four main modules of modally-related areas: Visual, Auditory,
  Somatosensory-Motor and Fronto-Limbic.}
\label{fig:CatNet}
\end{figure}

 
In this work we study the routes to synchronization in the
corticocortical network of cats brain (see Figure~\ref{fig:CatNet}) by
modelling each cortical area as a phase oscillator with an independent
internal frequency. This assumption considers that the ensemble of
neurons contained within a cortical area behaves coherently having a
well defined phase average whose dynamics is described by the internal
frequency~\cite{syncensembles}. The coupling between areas is modelled
using the Kuramoto nonlinear coupling and its relative strength can be
conveniently tuned to allow for the observation of synchronization at
different scales of organization. This seemingly crude approximation
allows to obtain similar synchronization patterns as those observed
with more refined models based on neuronal ensembles placed at each
cortical area~\cite{Zemanova1, Zemanova2, Zemanova3}. For instance,
using the Kuramoto model one obtains that highly connected areas
promote synchronization of neural activity just as revealed by the
more stylized model used for the dentate gyrus~\cite{Morgan}.

Our results point out that complex structures of highly connected
areas play a key role in the synchronization transition. In contrast
to the usual partition of the brain cortex into four sets of
modally-related areas, we find that this modular organization plays a
secondary role in the emergence of synchronization patterns. On the
contrary, we unveil that a new module made up of highly connected
areas (not necessarily modally-related) drives the dynamical
organization of the system. This new set is seen as the Rich-Club of
the corticocortical network. Surprisingly, the new partition of the
network including the Rich-Club as a module preserves
the modular behavior of the system's dynamics at low intercortical
coupling strengths. This modular behavior transforms into a
centralized one (driven by the Rich-Club) at the onset of global
synchronization highlighting the plasticity of the network to perform
specialized (modular) or integration (global) tasks depending on the
coupling scale.

\section*{Results}
As introduced above, we describe the dynamical behavior of the
cortical network using the Kuramoto model \cite{Kuramoto}, where the
time evolution of the phase of each cortical area, $\theta_{i}(t)$, is
given by
\begin{equation}
\dot{\theta_i}=\omega_i+\lambda\sum_{j=1}^{N}W_{ij}\sin(\theta_j-\theta_i),\;
\label{eq:1}
\end{equation}
where $\omega_{i}$ is the internal frequency associated to area $i$
and $W_{ij}$ is the weighted inter-cortical coupling matrix that takes
a value $0$ if area $j$ does not interact with the dynamics of area
$i$ while $W_{ij}>0$ otherwise. In this latter case $W_{ij}$ can take
values $1$, $2$ or $3$ depending on the axon density going from area
$j$ to area $i$. Let us remark that the inter-cortical coupling matrix
is not symmetric so that in general $W_{ij}\neq W_{ji}$. Besides, the
inter-cortical dynamical coupling is modeled as the sine of the phase
differences between two connected areas such that when
$\theta_{j}>\theta_{i}$ the average phase of area $i$ accelerates
while that of area $j$ slows down to approach each other. Finally, the
parameter $\lambda$ accounts for the strength of the inter-cortical
coupling.

In a system composed of all-to-all coupled oscillators, the Kuramoto
model shows a transition from incoherent dynamics to a synchronized
regime as $\lambda$ increases \cite{Kurarev1,Kurarev2}. However, when
the system has a nontrivial underlying structure this transition does
not take place in an homogeneous manner. In complex topologies, and
for moderate coupling values, certain parts of the system become
synchronized rather fast whereas other regions still behave
incoherently.  Therefore, one can monitor the synchronization patterns
that appear as the coupling $\lambda$ is increased and describe the
path to synchronization accurately\cite{SyncRev} by reconstructing the
synchronized subgraph composed of those nodes and links that share the
largest degree of synchronization (see Materials and Methods). The
study of these synchronization clusters as the coupling $\lambda$ is
increased allow to unveil the important set of nodes that drives these dynamical processes in the system.

We will analyze different scales of organization: $i)$ the macroscopic
scale referring to global synchronization of the network; $ii)$ the
microscopic scale of organization corresponding to the individual
state of the oscillators; and finally
$iii)$ the intermediate mesoscopic scale of dynamical organization
between the macroscopic and microscopic scales.  Usually, it consists
of groups of nodes classified according to a certain additional
information, for example that provided by the anatomico-functional
modules.  Every scale of organization requires a particular set of
statistical descriptors. This is especially important in the
mesoscopic scale where changing the groups, the characterization of
the system is also changed.

\subsection*{Macroscopic analysis}
We start by describing how global synchronization is attained as the
inter-cortical coupling $\lambda$ is increased. Global synchronization
is characterized by the usual Kuramoto order parameter, $r$, and the
fraction of links that are synchronized $r_{link}$ \cite{PRL,PRE} (see
Materials and Methods). Both parameters take values in the region
$[0,1]$, being close to $0$ when no dynamical coherence is observed
and close to $1$ when the system approaches to full
synchronization.

\begin{figure}[!ht]
\begin{center}
\includegraphics[width=4in]{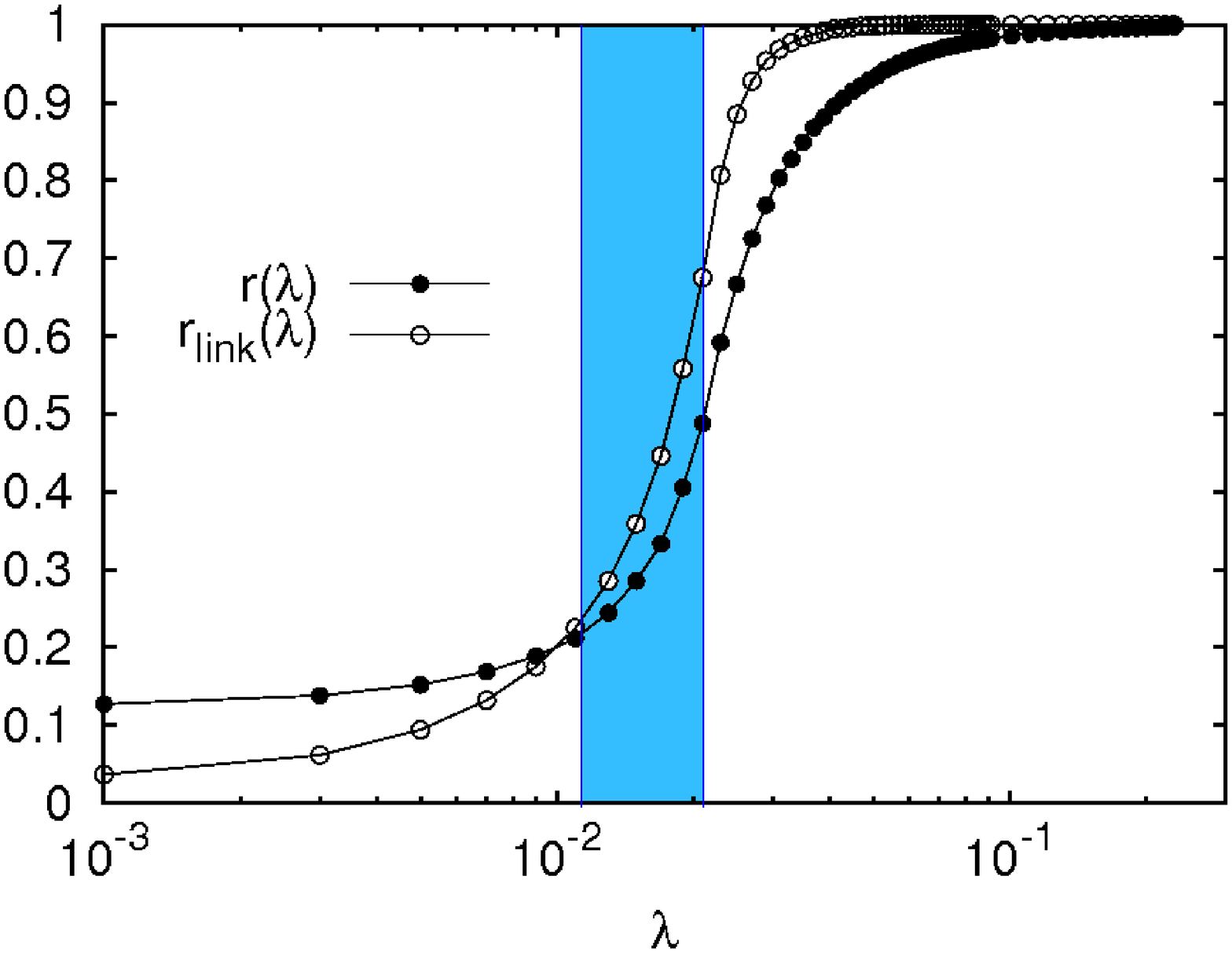}
\end{center}
\caption{ {\bf Synchronization diagrams.} The figure shows the
  evolution of the Kuramoto order parameter $r$ and the fraction of
  synchronized links $r_{link}$ as the coupling strenght is
  increased. The transition from asynchronous dynamics to global
  dynamical coherence as $\lambda$ grows is clear from the two
  curves. The region in blue corresponds to the onset of
  synchronization.}
\label{fig:diagram}
\end{figure}

In Fig.~\ref{fig:diagram} we show the evolution of $r$ and $r_{link}$
as a function of $\lambda$. The plot reveals a well defined transition
from incoherent to globally synchronized, the onset of synchronization
occurring for coupling strength $0.011\leq \lambda\leq 0.021$. When
$\lambda\simeq 0.2$, the system reaches the fully synchronized
state. In the following, we will explore this transition in more
detail and at lower scales of dynamical organization.

\subsection*{Mesoscopic analysis as described by the four anatomical modules}
The measures $r$ and $r_{link}$ describe completely the dynamical
state of the system if one assumes that all the cortical areas behave
identically. However, it is possible to extract more
information about the local dynamical properties of the system. In
particular, for a given value of $\lambda$ we can monitor the degree
of synchronization between two given areas $i$ and $j$,
  $r_{ij}\in [ 0,1]$ (see Materials and Methods).

The studies of the transition to synchronization in modular
architectures \cite{Alex-syncom,PRE} show that synchrony patterns
appear first at internal modules, {\em i.e.} synchrony shows up among
the nodes that belong to the same module due to a larger local density
within the module and similar pattern of inputs of the nodes.  As the
coupling $\lambda$ is increased, synchrony starts to affect the links
connecting nodes of different clusters and finally spreads to the
entire system.
Now, we analyze whether the four anatomico-functional modules of the
corticocortical network of the cat act also as dynamical clusters in
the synchronization transition. To this end, we have analyzed the
evolution of the average synchronization within and between the four
anatomical modules taking into account solely the information about
the dynamical coherence $r_{ij}$ between the network's areas. We
define the average synchronization between module $\alpha$ and module
$\beta$ as:
\begin{equation}
r_{\alpha\beta}=\frac{1}{L_{\alpha\beta}}\sum_{i\in\alpha,\;j\in\beta}r_{ij}\;,
\label{eq:intercluster}
\end{equation}
where $L_{\alpha\beta}$ is the number of possible pairs
  of areas from modules $\alpha$ and $\beta$, {\em i.e.},
  $L_{\alpha\beta}=N_{\alpha}N_{\beta}/2$ where $N_{\alpha}$ and
  $N_{\beta}$ are the number of areas of modules $\alpha$ and $\beta$
  respectively. If $\alpha=\beta$ equation (\ref{eq:intercluster})
denotes the intramodule average synchronization.

\begin{figure}[!ht]
\begin{center}
\includegraphics[width=6in]{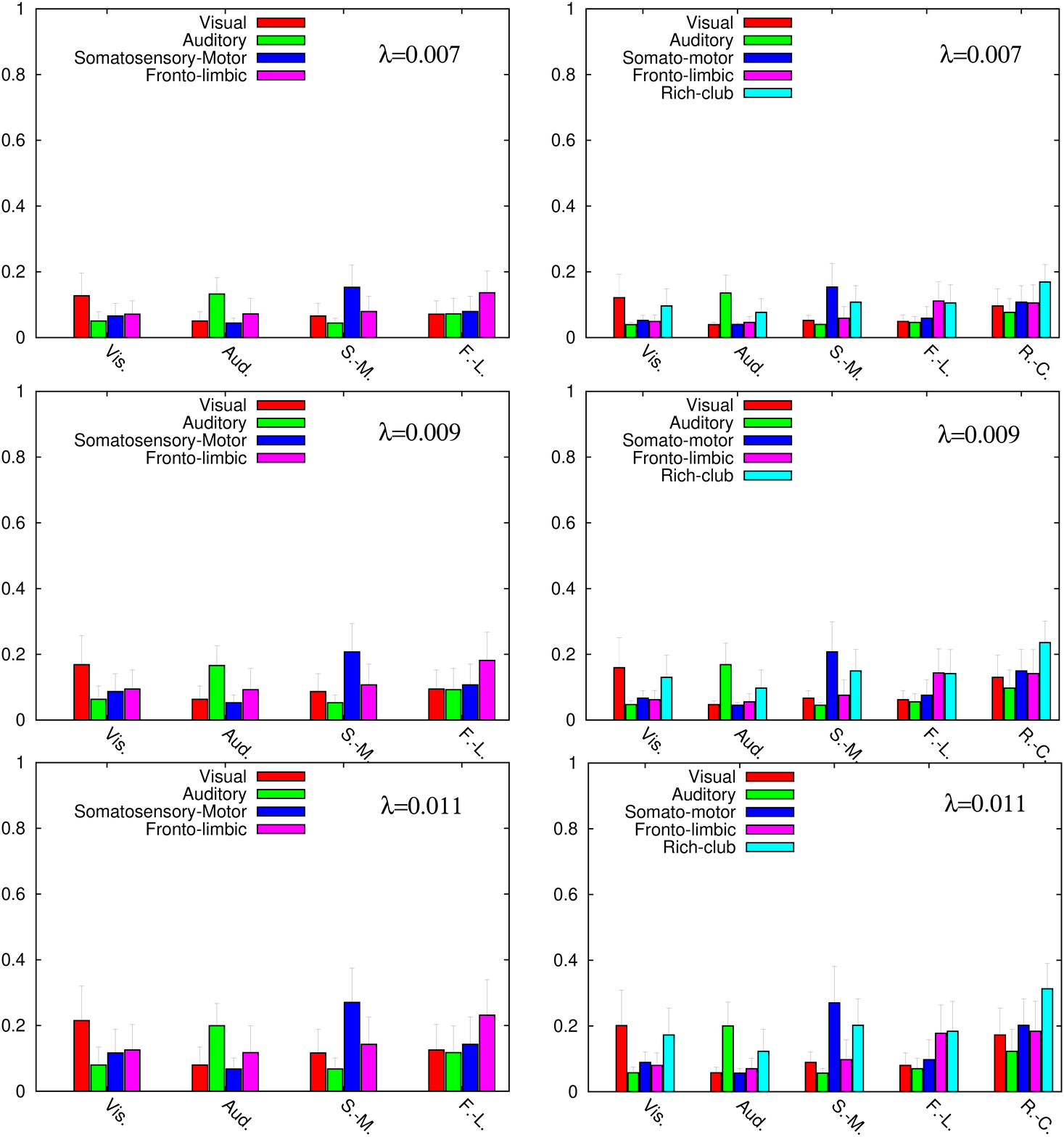}
\end{center}
\caption{ {\bf Dynamical correlation within the $4$ modal clusters
    (left column) and the new $4$ modally-related clusters and the
    Rich-Club (right column) before the onset of synchronization.} The
  bars of the histograms show the values of the dynamical correlation
  $r_{\alpha\beta}$ (see Equation (\ref{eq:intercluster})) between the
  $4$ original modules (left) and the new $4$ clusters and the
  Rich-Club (right). From top to bottom we show the cases for
  $\lambda=0.007$, $0.009$ and $0.011$ that correspond to the region
  before the onset of synchronization.  }
\label{fig:histo1}
\end{figure}

\begin{figure}[!ht]
\begin{center}
\includegraphics[width=5in]{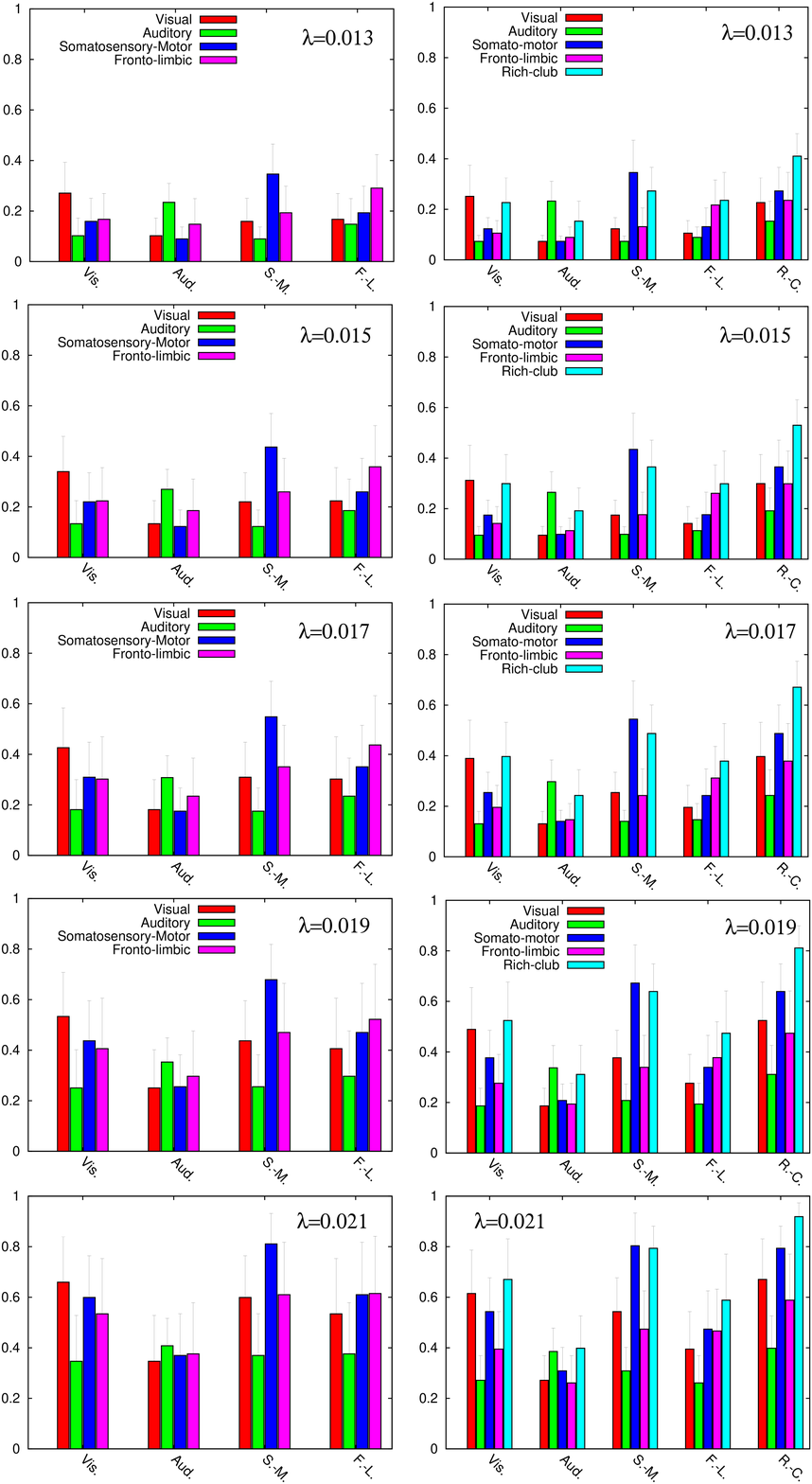}
\end{center}
\caption{ {\bf Dynamical correlation within the $4$ modal
    clusters (left column) and the $5$ dynamical clusters (right
    column) at the onset of synchronization.} The bars of the
  histograms show the values of the dynamical correlation
  $r_{\alpha\beta}$ (see Equation (\ref{eq:intercluster})) between the
  $4$ modules (left) and the new $4$ modally-related clusters and
  the Rich-Club (right). From top to bottom we show the cases for
  $\lambda=0.013$, $0.015$, $0.017$, $0.019$ and $0.021$ that
  correspond to the onset of synchronization.}
\label{fig:histo2}
\end{figure}

The histograms in the left columns of Figure~\ref{fig:histo1} and
Figure~\ref{fig:histo2} show the values of the set $\{r_{\alpha\beta}
\}$ for several values of $\lambda$ corresponding to the region before
(Fig.~\ref{fig:histo1}) and at (Fig.~\ref{fig:histo2}) the onset of
synchronization. From the histograms it is clear that the degree of
synchronization grows with $\lambda$ as it occurs for the global
parameters $r$ and $r_{link}$ in the macroscopic description. Besides,
the histograms inform us about the importance of the anatomical
partition in the dynamical organization of the cat cortex. From
Figure~\ref{fig:histo1} it becomes clear that the average dynamical
correlation within areas of the same anatomical module is higher than that between areas belonging to different
modules. Moreover, before the onset of synchronization,
for $\lambda=0.007$ to $\lambda=0.11$ all the modules
satisfy $r_{\alpha\alpha}\geq r_{\alpha\beta}$ for
$\alpha\neq\beta$. 
%
At the onset of synchronization (left column of
  Figure~\ref{fig:histo2}) we observe that the initial intra-module
  synchronization is progressively compensated by the increase of
  inter-module dynamical coherence. In particular, the influence of
  the Somatosensory-Motor on the remaining modules is remarkably
  relevant during the onset of synchronization and, for
  $\lambda>0.021$, this module shows the largest degree of
  synchronization with the rest of the system.

\subsection*{Microscopic analysis: Unveiling the dynamical organization}

The mesoscopic analysis based on the partition of the cortex into four
modules has revealed a fingerprint of a hierarchical organization of
the synchronization based on the dominating role of the
  Somatosensory-Motor module. Here we will analyze the microscopic
correlation between all the areas of the cortical network to unveil
whether there is a group of nodes that lead the onset of
synchronization in the system. To this purpose we study the subgraphs
formed by those pairs of areas sharing an average
synchronization value $r_{ij}$ larger than a threshold $T$. Certainly
when $T=1$ the subgraph is the null (empty) graph and for $T=0$ the subgraph is the whole cortex.

\begin{figure}[!ht]
\begin{center}
\includegraphics[width=6.2in]{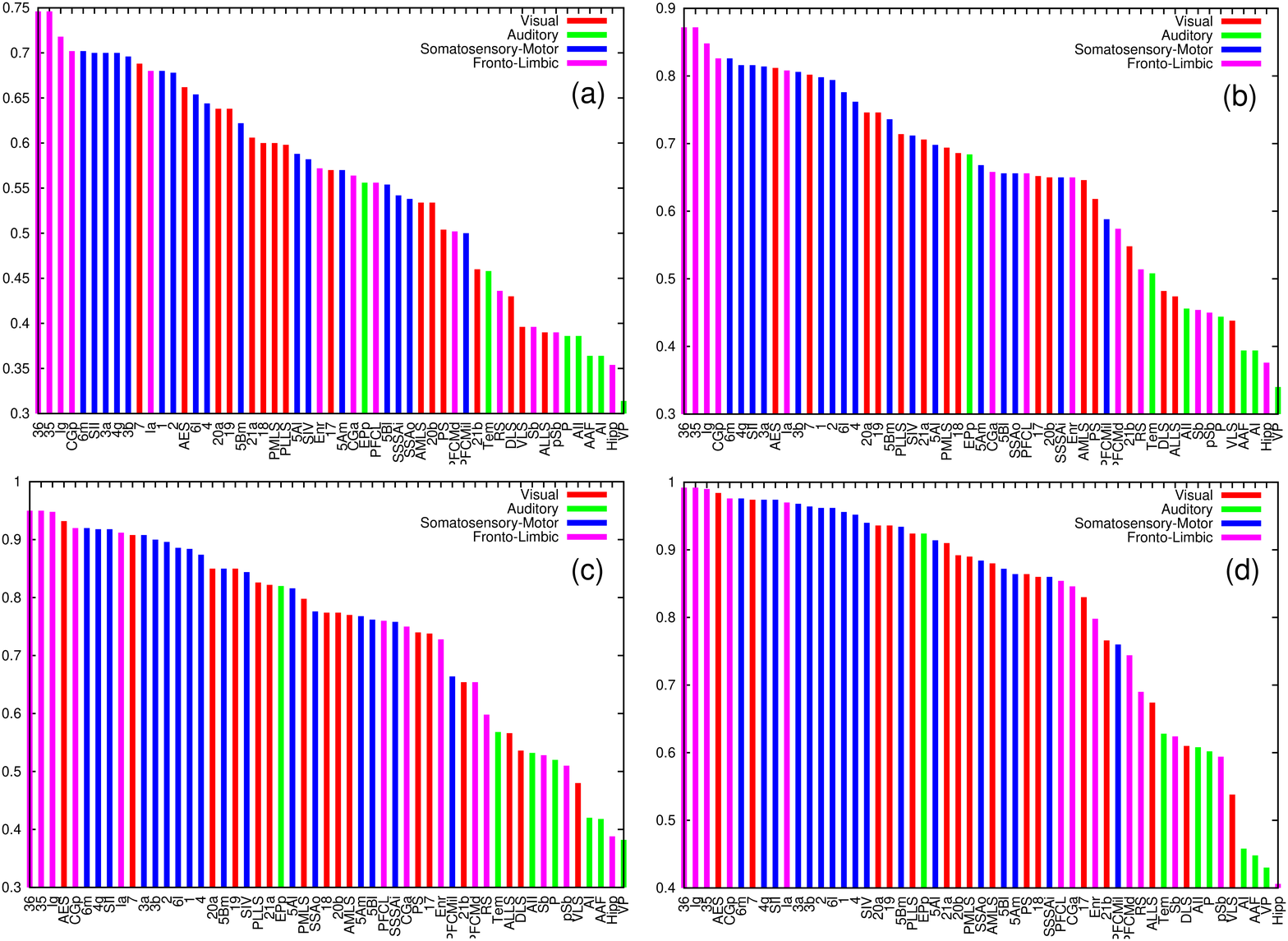}
\end{center}
\caption{{\bf Synchrony Rank of areas: Unveiling anatomical structure
    of the largest synchronized areas.} In these plots we show the
  rank of areas from the most to the less synchronized for
  $\lambda=0.015$ (a), $0.017$ (b), $0.019$ (c) and $0.021$ (d). The
  height of the bars account of the maximum value of the threshold,
  $T_{i}$, at which the area is incorporated in the synchronized
  subgraph. Besides, the colour of each bar accounts for the corresponding
  module of the cortical area.}
\label{fig:rank}
\end{figure}

In Figure \ref{fig:rank} we show a ranking of the cortical areas at
coupling strengths $\lambda=0.015$, $0.017$, $0.019$ and $0.021$
corresponding to the onset of synchronization. The rankings are made
by labeling the area $i$ with the largest value of the threshold at
which the area is incorporated into the synchronized subgraph as $T$
is tuned from $1$ to $0$. Additionally, the modular origin of the
areas has been color coded to distinguish the role of each module.
From the rankings we find that there are three areas {\em 36}, {\em
  35} and {\em Ig}, from the Fronto-Limbic system, that share the
largest degree of synchrony. In all the cases, several jumps in the
threshold are observed that distinguishes those groups of cortical
areas that are more synchronized than the rest of the network. For
instance, at $\lambda=0.017$ we observe $14$ areas spanning from the
{\em 36} area to the {\em 2} (Somatosensory-Motor system) while for
$\lambda=0.021$ we find $16$ areas ranging from the {\em 36} to the
area {\em 4} (Somatosensory-Motor system). From these figures it is
no clear that there is one module dominating the
synchronization. Quite on the contrary both Somatosensory-Motor and
Fronto-Limbic systems are well represented among the most synchronized
areas.

\begin{figure}[!ht]
\begin{center}
\includegraphics[width=6.2in]{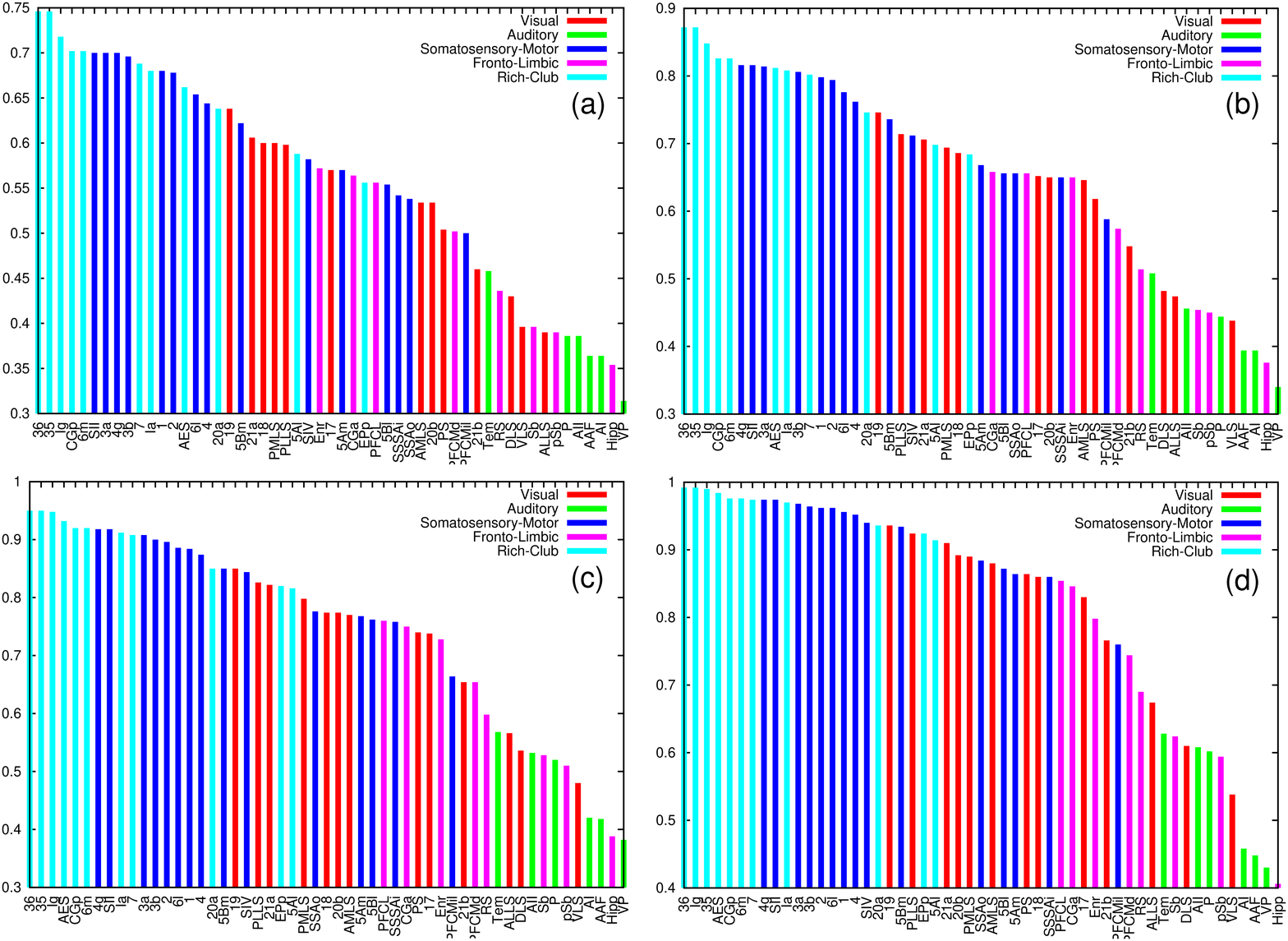}
\end{center}
\caption{{\bf Synchrony Rank of areas: Structure of the largest
    synchronized areas as described by the Rich-Club.} The two plots
  show the same synchrony ranks as in Figure \ref{fig:rank} (again
  $\lambda=0.015$ (a), $0.017$ (b), $0.019$ (c) and $0.021$ (d)). We
  have recolored the bars of those areas corresponding to the
  Rich-Club to highlight the dominant role of this topological
  structure in the composition of the largest synchronized areas.}
\label{fig:newrank}
\end{figure}

A further analysis of the composition of these highly synchronized
areas reveals that most of them take part in a higher-order
topological structure of the cortical network: a Rich-Club (see
Materials and Methods). The Rich-Club of a given network is made up of
a set of nodes with high connectivity, which at the same time, form a
tightly interconnected community~\cite{Zhou_RichClub, RichClub}. Therefore, the
Rich-Club of a network can be described as a highly cohesive set of
hubs, that form a dominant community in the hierarchical
organization. The Rich-Club of the cat cortex is composed of $11$
cortical areas of different modalities: $3$ visual areas ({\em 20a},
{\em 7} and {\em AES}), $1$ area from the Auditory system ({\em EPp}),
$2$ areas of the Somatosensory-Motor system ({\em 6m} and {\em 5Al})
and $5$ fronto-limbic areas ({\em Ia}, {\em Ig}, {\em CGp}, {\em 35}
and {\em 36}). In Figure \ref{fig:newrank} we show again the ranking
of areas for the cases $\lambda=0.015$, $0.017$, $0.019$ and $0.021$
but highlighting those areas belonging to the Rich-Club in black. From
the plots it is clear that most of the Rich-Club areas are largely
synchronized.
In particular for $\lambda=0.019$ and $0.021$ the $8$ out of the $10$
largest synchronized areas of the network belong to the Rich-Club,
although originally they belong to the Fronto-Limbic,
Somatosensory-Motor and the Visual systems in the partition into four
modules.

\subsection*{Mesoscopic analysis of synchronization including the Rich-Club}

Looking at the composition of the Rich-Club we observe that it is
mainly composed of fronto-limbic areas. Taking into account that we
previously observed how the Somatosensory-Motor system took the
leading role within the description with $4$ modules of the
synchronization transition, this dominance of the Fronto-Limbic system
in the Rich-Club may seem counterintuitive. To test the role of the
Rich-Club in the synchronization transition we define a new partition
of the cortical network into $5$ clusters composed of the Rich-Club
(as defined above) and the $4$ original modules, but with the
corresponding areas of the Rich-Club removed from them.

At the mesoscopic scale, we investigate the self-correlation of the
new five clusters and their cross correlation according to Equation
(\ref{eq:intercluster}). In the right columns of Figure
\ref{fig:histo1} and Figure \ref{fig:histo2} we present the histograms
of the inter and intra correlations for different values of the
coupling.
In both figures the role of the Rich-Club orchestrating the process
towards synchrony while increasing the coupling strength becomes
clear.
More importantly, the addition of the Rich-Club to the partition helps
to elucidate the patterns of synchrony: both the dynamical
self-correlation of the four modally-related clusters and their correlation
with the Rich-Club remain large.
In particular, we observe in Figure \ref{fig:histo1} that, before the
onset of synchronization, these new five modules keep a large
self-correlation during this stage. On the other hand, at the onset of
synchronization (Figure \ref{fig:histo2}) the four modally-related
clusters loose their large self-correlation in the following sequence:
The ``Fronto-Limbic'' cluster remains autocorrelated until
$\lambda=0.013$, the ``Visual'' one until $\lambda=0.017$, the
``Auditory'' system until 0.019 and the ``Somatosensory-Motor''
cluster until $\lambda=0.021$. For larger couplings, all the clusters
switch from autocorrelation to be synchronized with the Rich-Club,
which acts as a physical mean-field of the system. Moreover, during
the whole synchronization path the Rich-Club is always the cluster
with the largest self-correlation.
Thus, the distinction of Rich-Club in the partition preserves the autocorrelation of the four modal clusters before the
onset of synchronization while, at the same time, rules the path to
complete dynamical coherence during the onset of synchronization.
This two-stage dynamics (modal cluster synchronization followed
by a sequential synchronization with the Rich-Club) supports the idea
that the modular organization with a centralized hierarchy described
in~\cite{ZamoraIntegrator} facilitates the segregation and integration
of information in the cortex.

\subsection*{Characterization of the transition from modular to centralized synchronization}

The results so far indicate a plausible transition from modular to
centralized organization in the cortex, depending on the coupling
strength. In particular, we have shown patterns of synchronization
that change the behavior while increasing the coupling $\lambda$. Now
we propose a characterization of this change in terms of statistical
descriptors. To this end, we define two different measures: {\em (i)}
the dynamical modularity (DM) and {\em (ii)} the dynamical
centralization (DC). The dynamical modularity compares the degree of
internal synchrony within the clusters with the average dynamical
correlation across clusters. With this aim we define the DM as the
fraction of the average self-correlation of clusters and the average
intercluster cross-correlation. For a network composed of $m$ clusters
we have:
\begin{equation}
DM=\frac{\sum_{\alpha}r_{\alpha\alpha}/m}{\sum_{\alpha,\beta\neq\alpha}r_{\alpha\beta}/[m(m-1)]}\;.
\end{equation}
The DM will take values above $1$ when the system contains true
dynamical clusters while $DM<1$ means that the entitity of the
partition is not consistent with a clustered behavior. On the other
hand, the dynamical centralization of the network measures the
relative difference in synchrony between the maximum among the $m$
clusters of $r_{\alpha}= \sum_{\beta} r_{\alpha\beta}/m$ and the
average degree of synchrony over clusters, $\langle
r_{\alpha}\rangle=\sum_{\alpha}r_{\alpha}/m$:
\begin{equation}
DC= \frac{\max_{\alpha}\{r_{\alpha}\} -<r_{\alpha}>}{<r_{\alpha}>}\;.
\label{eq:DC}
\end{equation}
In the case of the DC we always obtain positive values. A large value
of DC means that the system displays a highly centralized dynamical
behavior around a leading cluster while we will obtain $DC$ values
approaching to $0$ when the system behaves homogeneously, {\em i.e.}
when there is no leading cluster that centralizes the dynamics.

\begin{figure}[!ht]
\begin{center}
\includegraphics[width=4in]{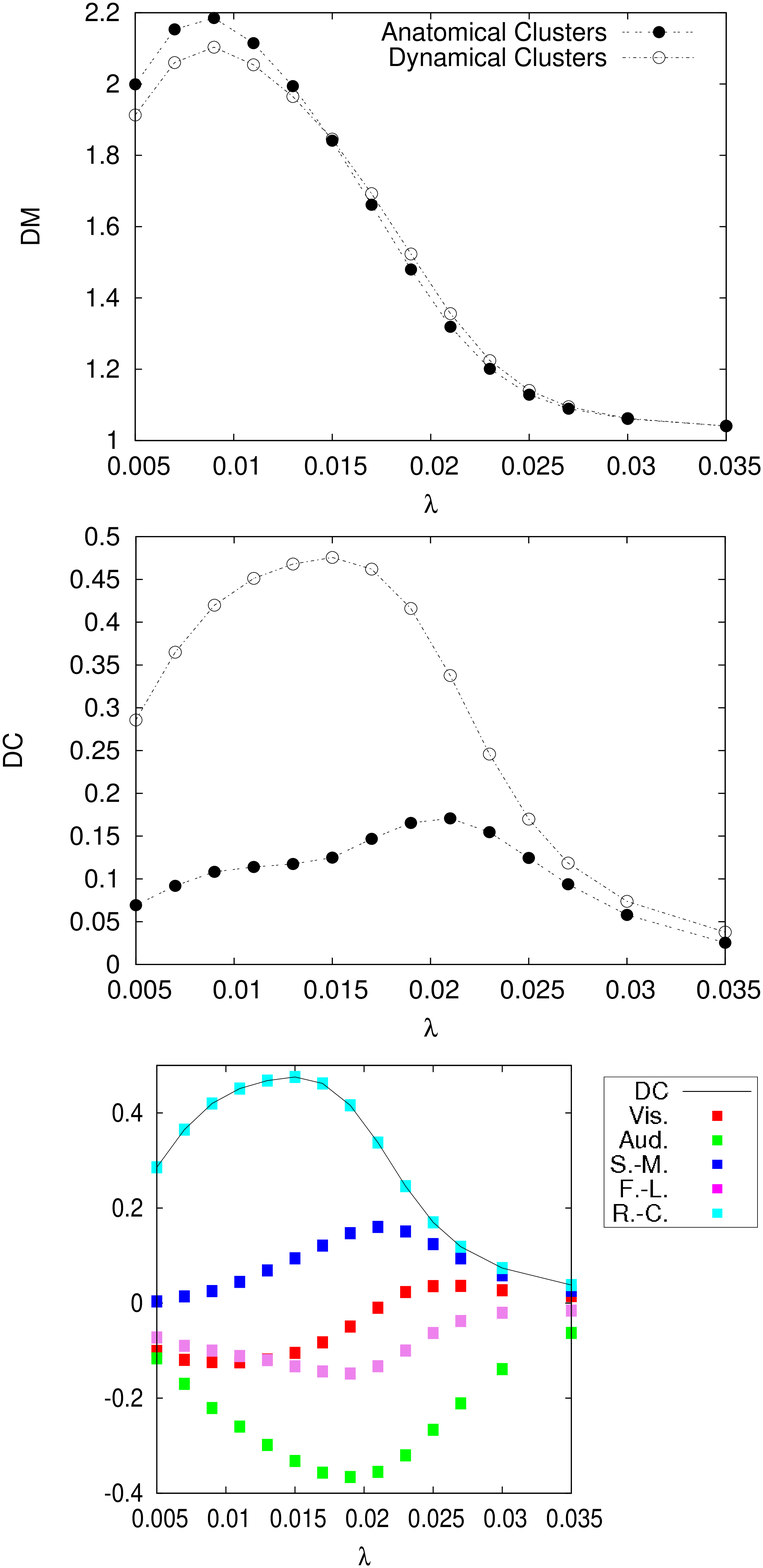}
\end{center}
\caption{{\bf Transition from Modular to Centralized organization of
    synchronization.} The two upper plots show the evolution of the
  dynamical modularity $DM$ and the dynamical centralization $DC$ as a
  function of $\lambda$. Both panels show the evolution of the above
  properties for the network described by means of both the original $4$
  modal clusters and the $5$ new modules including the
  Rich-Club. The panel in the bottom shows the evolution of the
  DC of the $5$ new modules obtained by replacing in Equation
  (\ref{eq:DC}) the term $\max_{\alpha}\{r_{\alpha}\}$ by each
  $r_{\alpha}$ corresponding to the new $5$ modules. From the curves
  it is clear that the dynamics is centralized around the Rich-Club.
\label{fig:orderparam}}
\end{figure}

We have measured both DM and DC for the original partition into $m=4$
modules and the new partition with $m=5$ incorporating the
Rich-Club. In Figure \ref{fig:orderparam} we show the evolution of the
two quantities as a function of the coupling parameter. For the case
of the DM we confirm that the partition with the Rich-Club keeps the
modular behavior of the original partition along the whole
synchronization path. The DM is remarkably high for low values of the
coupling $\lambda$ pointing out that before the synchronization onset
the internal synchronization dominates over the cross-correlation
between the clusters. Regarding the DC we find that in both partitions
$DC$ increases with $\lambda$ reaching a maximum around the
synchronization onset, signaling that at this point the
synchronization is driven hierarchically and lead by one of the
modules. However, the partition that incorporates the Rich-Club shows
the remarkably larger values of DC along the whole path, specially
around the synchronization onset. In particular, the dominant role of
the Rich-Club is clearly highlighted by the maximum of the DC at
$\lambda=0.015$, just at the onset of synchronization. In order to
verify that the Rich-Club is the cluster contributing to the term
$\max_{\alpha}\{r_{\alpha}\}$ in the dynamical centrality of the
network we plot in Figure \ref{fig:orderparam} the evolution of the DC
considering each of the $5$ modules as the the central cluster by
substituting $\max_{\alpha}\{r_{\alpha}\}$ by the corresponding value
of $r_{\alpha}$. From the plot it is clear that the Rich-Club is the
central cluster orchestrating the dynamics of the system at the onset
of synchronization.

The coupled evolution of DM and DC corroborates the two-mode operation 
of the cortical network when described with the Rich-Club and the 
remaining parts of the four original modules: At low values 
of the coupling, the modular structure of the network dominates the 
synchronization dynamics, pointing out the capacity to concentrate 
sensory stimuli within its corresponding module. When the coupling 
is increased the dynamical organization is 
driven by a leading subset of nodes, organized in a topological
Rich-Club, that integrates information between different regions of
the cortex.

\section*{Discussion}

Previous simulations performed in the cat cortical network
\cite{Zemanova1,Zemanova2,Zemanova3} have dealt with its
synchronization properties. In these works, the transition towards
synchronization is studied by using ensembles of neurons coupled
through a small-world topology placed inside each cortical area
whereas different neuronal populations are dynamically coupled
accordingly to the topology of the cat cortical network. By means of
this two-level dynamical model, numerical simulations allowed to find
different clusters of synchrony as the coupling between the cortical
areas is increased. It was found that only for weak coupling these
clusters were closely related to the four modal clusters.
In the light of these previous studies, and the recent report of a
novel modular and hierarchical organization of the corticocortical
connectivity~\cite{ZamoraIntegrator}, the issue regarding the relation
between the mesoscopic structure of the cat cortex and its dynamical
organization remains open.

Here, we have investigated the evolution of synchronization in a
network representing the actual connectivity among cortical areas in
the cat's brain.  We have confirmed, that the role of the different
areas in the path towards synchrony is difficult to assess using the
traditional partition into four groups of modally-related areas. On
the contrary, we have shown that a subset of areas, forming a
topological Rich-Club, orchestrates this process. The distinction of
this subset permits the interpretation of a new mesoscale formed by
the four modules, excluding some nodes that form the Rich-Club, which
are considered here as the fifth module. This proposed structure
allows us to reveal a transition in the path to synchronization as a
function of the coupling strength, that seems to indicate a two-mode
operation strategy. For low values of the coupling, a state of weak
internal coherence within the five modules governs the coordination
dynamics of the network. As the coupling strength is increased, the
Rich-Club becomes the responsible of centralizing the network dynamics
and leads the transition towards global synchronization.


Finally, the composition of the Rich-Club allows to make some
additional biologically relevant observations. First, the Rich-Club
comprises of cortical areas of the four different modalities,
supporting the hypothesis of distributed coordination dynamics at the
highest levels of cortical processing such as integration of
multisensory information~\cite{Bressler2001}.  Second, the Rich-Club
comprises of most of the frontal areas in the Fronto-Limbic
module. Moreover, the areas of the Rich-Club collected from the
original Somatosensory-Motor system contain the so-called
supplementary motor area (SMA). The SMA is a controversial region of
the motor cortex, since in contrast with the rest of
somatosensory-motor areas it is in charge of the initiation of planned
or programmed movements \cite{SMS}. Furthermore, the area {\em AES} of
the Rich-Club, originally assigned to the Visual module, is believed
to integrate all visual and even auditory signals for their multimodal
processing and transference as coherent communication signals
\cite{AES1}. Summing up, the Rich-Club is basically made up of areas
involved in higher cognitive tasks devoted to planning and
integration.
The prominent role of the aforementioned regions in the cortex
activity is unveiled from our network perspective in terms of a
Rich-Club leading the path to synchronization. Our proposal, after
this observation, is to investigate the evolution of synchronization
in the cat cortex by tracking the transient of five modules
corresponding to the anatomico-functional areas (S-M, F-L, Aud, Vis)
and the Rich-Club as a separate (but interrelated) functional
entities.

\section*{Materials and Methods}


\subsection*{Cortico-cortical network of cats' brain}

After an extensive collation of literature reporting anatomical
tract-tracing experiments, Scannell and Young~\cite{Scannell1993,
Scannell1995} published a dataset containing the corticocortical and
cortico-thalamical projections between regions of one brain hemisphere
in cats. The connections were weighted according to the axonal density
of the projections. Connections originally reported as \emph{weak} or
\emph{sparse} were classified with 1 and, the connections originally
reported as \emph{strong} or \emph{dense} with 3. The connections
reported as \emph{intermediate} strength, as well as those connections
for which no strength information was available, were classified with
2, see Figure~\ref{fig:CatNet}(b). Here we make use of a version 
of the network~\cite{Hilgetag2000} consisting of $N = 53$ cortical areas 
interconnected by $L = 826$ directed corticocortical projections.

\subsection*{Rich-Club areas}

A key factor of the hierarchical organization of the corticocortical
network of the cat is that the hub areas (those with the largest
number of projections) are very densely connected between
them~\cite{ZamoraIntegrator}. The Rich-Club
phenomenon~\cite{Zhou_RichClub,RichClub} is characterised by the
growth of the internal density of links between all nodes with degree
larger than a given $k'$, referred as $k$-density, $\phi(k')$:
\begin{equation}
\phi(k') = \frac{L_{k'}}{N_{k'} (N_{k'} -1)},
\end{equation}
where $N_{k'}$ is the number of nodes with $k(v) \geq k'$ and $L_{k'}$
is the number of links between them. As $\phi(k)$ is an increasing
function of $k$, a conclusive interpretation requires the comparison
with random surrogate networks with the same degree distribution. The
question is then whether $\phi(k)$ of the real network grows faster or
slower with $k$ than the expected $k$-density of the surrogate
networks. If $\phi(k)$ grows slower, it means that
the hubs are more independent of each other than expected.

\begin{figure}
\begin{center}
\includegraphics[width=0.80 \textwidth,clip=]{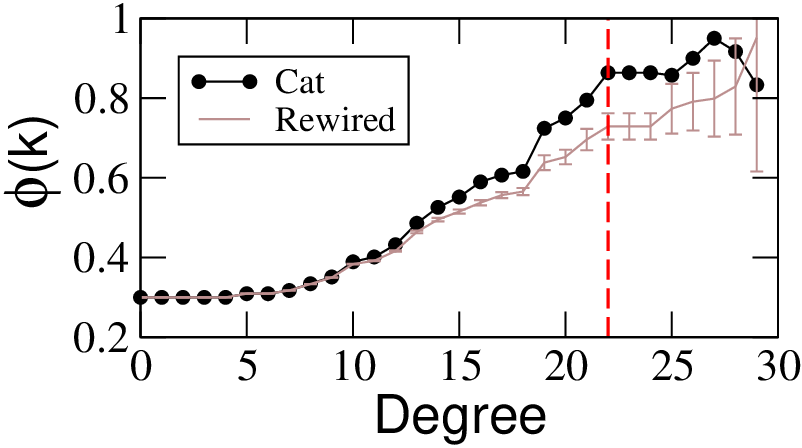}
\end{center}
\caption{K-density of the corticocortical directed network of the cat
  $\phi(k)$, compared to the expectation out of the surrogate
  ensemble. The largest difference occurs at $k=22$ (vertically dashed
  line) giving rise to a Rich-Club composed of eleven cortical
  areas.
  \label{fig:RichClub}}
\end{figure}

In our case, the network is directed but the input degree $k_{in}(v)$
and the output degree $k_{out}(v)$ of the areas are highly
correlated. Hence, we compute the $k$-density of the corticocortical
network of the cat, $\phi_{cat}(k)$, considering the degree of every
area as: $k(v) = \frac{1}{2}(k_{in}(v) + k_{out}(v))$.  The result is
presented in Figure~\ref{fig:RichClub} together with the ensemble
average $\phi_{1n}(k)$ of $100$ surrogate networks. At low degrees
$\phi_{cat}(k)$ follows very close the expectation, but for degrees $k
> 13$, $\phi_{cat}(k)$ starts to grow faster. The largest difference
occurs at $k = 22$, comprising of a set of eleven cortical hubs of the
four modalities: visual areas {\em 20a}, {\em 7} and {\em AES};
auditory area {\em EPp}, somatosensory-motor areas {\em 6m} and {\em
  5Al}; and fronto-limbic areas {\em Ia}, {\em Ig}, {\em CGp}, {\em
  CGa}, {\em 35} and {\em 36}. \\

\subsection*{Numerical simulation details}

We integrate the Kuramoto equations, see equation (\ref{eq:1}), using
a fourth order Runge-Kutta method with time step $\delta
t=10^{-2}$. The system is set up by randomly assigning the initial
conditions $\{\theta_{i}(0)\}$ and the internal frequencies
$\{\omega_{i}\}$ randomly in the intervals $[-\pi,\pi]$ and
$[-1/2,1/2]$ respectively. The integration of the Kuramoto is
performed for a total time $T=700$. After a transient time of
$\tau=300$ we start the computation of the different dynamical
measures such as the order parameters $r$ and $r_{link}$.

\subsection*{Synchronization order parameters}

The dynamical coherence of the population of $N$ oscillators (areas)
is measured by means of two different order parameters $r$ and
$r_{link}$. The first one is obtained from a complex
number $z(t)$ defined as follows:
\begin{equation}
z(t)=r(t)\exp\left[{\mbox i}\phi(t)\right]=\sum_{j=1}^{N}\exp\left[{\mbox i}\theta_{j}(t)\right]\;.
\end{equation}
The modulus of $z(t)$, $r(t)$, measures the phase coherence of the
population while $\phi(t)$ is the average phase of the population of
oscillators. Averaging over time the value of $r(t)$ we obtain the
order parameter $r=\langle r(t)\rangle$.

The second order parameter, $r_{link}$, is measured looking at the
local synchronization patterns, allowing for the exploration of how
global synchronization is attained. We define $r_{link}$ by measuring
the degree of synchrony between two connected areas $i$ and $j$:
\begin{equation}
C_{ij}=\lim_{\Delta t\rightarrow\infty}\left|\frac{A_{ij}}{\Delta
  t}\int_{\tau}^{\tau+\Delta t}{\mbox e}^{{\mbox
    i}\left[\theta_{i}(t)-\theta_{j}(t)\right]}\right|\;,
\label{eq:cij}
\end{equation}
where $A_{ij}$ is the adjacency matrix of the network, being
$A_{ij}=1$ when $W_{ij}>0$ and $A_{ij}=0$ otherwise.  Each of the
values $\{C_{ij}\}$ are bounded in the interval $[0,1]$, being
$C_{ij}=1$ when the connected areas $i$ and $j$ are fully synchronized
and $r_{ij}=0$ when these areas are dynamically uncorrelated. Note
that for a correct computation of $C_{ij}$ the averaging time $\Delta
t$ should be taken large enough (in our computations $\Delta t=400$)
in order to obtain good measures of the degree of coherence between
each pair of areas. Since $C_{ij}=0$ for the areas that are not
physically connected we construct the $N\times N$ matrix $C$ and
define the global order parameter $r_{link}$ as follows:
\begin{equation}
r_{link}=\frac{1}{L}\sum_{i,j}C_{ij}\;.
\label{eq:rlink}
\end{equation}
Therefore, the parameter $r_{link}$ measures the fraction of all
possible links that are synchronized in the network. 

In the more general case in which all possible pairs of areas are taken into account to compute the average synchronization between cortical areas, Eq.(\ref{eq:cij}) and Eq.(\ref{eq:rlink}) can be rewritten as:
\begin{equation}
C^{*}_{ij}=\lim_{\Delta t\rightarrow\infty}\left|\frac{1}{\Delta
  t}\int_{\tau}^{\tau+\Delta t}{\mbox e}^{{\mbox
    i}\left[\theta_{i}(t)-\theta_{j}(t)\right]}\right|\;,
\label{eq:cij2}
\end{equation}
and 
\begin{equation}
r^{*}_{link}=\frac{2}{N(N-1)}\sum_{i,j}C^{*}_{ij}\;,
\label{eq:rlink2}
\end{equation}
respectively. Note that $C^{*}_{ij}$ and $r^{*}_{link}$ account for the degree of synchronization between areas $i$ and $j$ regardless of whether or not they are connected.

\subsection*{Defining the average synchronization between areas}
To label two areas $i$ and $j$ as synchronized or not one has to
analyze the matrix $C^{*}$ and construct a filtered matrix $F$ whose
elements are either $F_{ij}=1$ if $i$ and $j$ are considered as
synchronized or $F_{ij}=0$ otherwise. From the computation of
$r^{*}_{link}$, equation (\ref{eq:rlink2}), one knows the fraction of
all possible pairs of areas that are synchronized. Therefore, one
would expect that $N(N-1)\cdot r^{*}_{link}/2$ elements of the matrix
$F$ have $F_{ij}=1$, while the remaining elements are $F_{ij}=0$. The
former elements correspond to the $N(N-1)\cdot r^{*}_{link}/2$ pairs
with the largest values of $C^{*}_{ij}$.

In order to measure the average degree of synchronization between
pairs of areas one have to average over different $n$ realizations
using different initial conditions $\{\theta_{i}(0)\}$ and different
internal frequencies $\{\omega_{i}\}$ (typically we have used
$n=5\cdot 10^3$ different realizations for each value of $\lambda$
studied). To this purpose we average the set of filtered matrices
$\{F^{l}\}$ ($l=1$,...,$n$) of the different realizations to obtain
the average degree of synchronization between areas:
\begin{equation}
r_{ij}=\frac{1}{n}\sum_{l=1}^{n}F_{ij}^l\;.
\end{equation}
In this way the value for $r_{ij}\in[0,1]$ accounts for the
probability that areas $i$ and $j$ are considered as synchronized.

\section*{Acknowledgments}
The authors are grateful to Jes\'us G\'omez-Tol\'on for useful
discussions and suggestions that helped to improve the manuscript.
%
%
%
%
%


\begin{thebibliography}{99}

\bibitem{Engel} Engel AK, Singer W (2001) Temporal binding and the
  neural correlates of sensory awareness. Trends Cogn. Sci. {\bf 5}:
  16--25.

\bibitem{EngelReview} Engel AK, Fries P, Singer W (2001) Rapid
  feature selective neuronal synchronization through correlated
  latency shifting. Nat. Rev. Neurosc. {\bf 2}: 704--716.

\bibitem{Fahle} Fahle M (1993) Figure--Ground Discrimination from
  Temporal Information. Proc. R. Soc. Lond. B {\bf 254}: 199--203.

\bibitem{Singer} Singer W, Gray CM (1995) Visual feature integration
  and the temporal correlation hypothesis. Ann. Rev. Neurosci. {\bf
    18}: 555--586.

\bibitem{Uhlhaas} Ulhaas PJ, Pipa G, Lima B, Melloni L, Neuenschwander
  S {\em et al.} (2009) Neural synchrony in cortical networks:
  history, concept and current status. Frontiers Int. Neurosc. {\bf
    3}, 17.

\bibitem{Melloni} Melloni L, Molina C, Pena M, Torres D, Singer W,
  {\em et al.} (2007) Synchronization of neural activity across
  cortical areas correlates with conscious perception. J. Neurosc {\bf
    27}: 2858 --2865.



\bibitem{Handbook} Boccaletti S., Latora V. and Moreno Y.(eds.)
  (2009), {\em Handbook on Biological Networks}, Singapore: World
  Scientific.

\bibitem{Scannell1995} Scannell JW, Blakemore CW, Young MP (1995)
  Analysis of connectivity in the cat cerebral
  cortex. J. Neurosc. {\bf 15}: 1463--1483.

\bibitem{Scannell1993} Scannell JW, Burns GAPC, Hilgetag CC, O'Neill
  MA, Young MP (1999) The connectional organization of the
  cortico-thalamic system of the cat. Cer. Cortex {\bf 9}: 277--299.

\bibitem{Hilgetag2000} Hilgetag CC, Burns GAPC, O'Neill MA, Scannell
  JW, Young MP (2000) Anatomical connectivity defines the organization
  of clusters of cortical areas in the macaque monkey and the
  cat. Phil. Trans. R. Soc. London B {\bf 355}: 91--110.

\bibitem{Hilgetag2} Hilgetag CC, O'Neill MA, Young MP (2000)
  Hierarchical organization of macaque and cat cortical sensory
  systems explored with a novel network
  processor. Phil. Trans. R. Soc. London B {\bf 355}: 71--89.

\bibitem{Sporns} Sporns O, Chialvo DR, Kaiser M, Hilgetag CC (2004)
  Organization, development and function of complex brain networks.
  Trends Cogn. Sci. {\bf 8}: 418--425.

\bibitem{Hilgetag2004} Hilgetag CC, Kaiser M (2004) Clustered organization
  of cortical connectivity. Neuroinformatics {\bf 2}: 353--360.

\bibitem{Zamora2009} Zamora-L\'opez G, Zhou CS, Kurths J (2009) Graphs
  analysis of cortical networks reveals complex anatomical
  communication substrate. Chaos {\bf 19}: 015117.

\bibitem{SpornsHubs} Sporns O, Honey CJ, K\"otter R (2007)
  Identification and classification of hubs in brain networks. PLoS
  ONE {\bf 10}: e1049.

\bibitem{ZamoraIntegrator} Zamora-L\'opez G., Zhou CS, Kurths J (2010)
  Cortical hubs form a module for multisensory integration on top of
  the hierarchy of cortical networks. Frontiers Neuroinf. {\it in
    press}.

\bibitem{HagmannHubs} Hagmann P, Cammoun L, Gigandet X, Meuli R, Honey
  CJ {\em et al.} (2008) Mapping the Structural Core of Human Cerebral
  Cortex. PLoS Biol {\bf 6(7)}: e159.


\bibitem{SyncRev} Arenas A, D\'{\i}az-Guilera A, Kurths J, Moreno Y,
  Zhou CS (2008) Synchronization in complex networks. Phys. Rep. {\bf
    469}: 93--153.

\bibitem{Alex-syncom} Arenas A, D\'{\i}az-Guilera A, P\'erez-Vicente
  CJ (2006) Synchronization reveals topological scales in complex
  networks. Phys. Rev. Lett. {\bf 96}: 114102.

\bibitem{PRL} G\'omez-Garde\~nes J, Moreno Y,Arenas A (2007) Paths to
  synchronization in complex networks. Phys. Rev. Lett.{\bf 98}:
  034101.

\bibitem{PRE} G\'omez-Garde\~nes J, Moreno Y, Arenas A (2007)
  Synchronizability determined by coupling strengths in complex
  networks. Phys. Rev. E {\bf 75}: 066106.

\bibitem{syncensembles} Rulkov NF (2001) Regularization of
  synchronized chaotic burst. Phys. Rev. Lett. {\bf 86}: 183.

\bibitem{Zemanova1} Zemanova L, Zhou CS, Kurths J (2006) Structural
  and functional clusters of complex brain networks. Physica D {\bf
    224}: 202--212.

\bibitem{Zemanova2} Zhou CS, Zemanova L, Zamora G, Hilgetag CC, Kurths
  J (2006) Hierarchical organization unveiled by functional
  connectivity in complex brain networks. Phys. Rev. Lett. {\bf 97}:
    238103.

\bibitem{Zemanova3} Zhou CS, Zemanova L, Zamora G, Hilgetag CC, Kurths
  J (2007) Structure-function relationship in complex brain networks
  expressed by hierarchical synchronization. New J. Phys. {\bf 9}:
  178.

\bibitem{Morgan} Morgan RJ, Soltesz I (2008), Nonrandom connectivity
  of the epileptic dentate gyrus predicts a major role for neuronal
  hubs in seizures. Proc. Natl. Acad. Sci. (USA) {\bf 105}: 6179--6184.

\bibitem{Kuramoto} Kuramoto Y (1984) Cooperative Dynamics of
  Oscillator Community. Prog. Theor. Phys. {\bf 79}: 223--240.

\bibitem{Kurarev1} Strogatz SH (2000) From Kuramoto to Crawford:
  exploring the onset of synchronization in populations of coupled
  oscillators. Physica D {\bf 143}: 1--20.

\bibitem{Kurarev2} Acebron JA, Bonilla LL, Perez Vicente CJ, Ritort F,
  Spigler R (2005) The Kuramoto model: A simple paradigm for
  synchronization phenomena. Rev. Mod. Phys. {\bf 77}: 137--185.

\bibitem{Zhou_RichClub} Zhou S and Modrag\'on RJ (2004) The rich-club 
phenomenon in the internet topology. IEEE Comm. Lett. {\bf 8}(3): 80--182.

\bibitem{RichClub} Colizza V, Flammini A, Serrano MA, Vespignani A
  (2006) Detecting Rich-Club ordering in complex networks. Nature
  Phys. {\bf 2}: 110-115.

\bibitem{Bressler2001} Bressler SL and Kelso JAS (2001) Cortical
  coordination dynamics and cognition.  Trends Cogn. Sci. {\bf 5},
  26--36.

\bibitem{SMS} Nachev P, Kennard Ch, Husain M (2008) Functional role of
  the supplementary and pre-supplementary motor
  areas. Nature Rev. Neuroscience {\bf 9}: 856--869.

\bibitem{AES1} Stein BE, Stanford TR (2008) Multisensory integration:
  current issues from the perspective of the single neuron. Nature
  Rev. Neuroscience {\bf 9}: 255--266.

\end{thebibliography}

\newpage


\section*{Figure Legends}



\end{document}